\documentclass[%
 reprint,
 amsmath,amssymb,
 aps,
]{revtex4-2}

\usepackage{graphicx}
\usepackage{dcolumn}
\usepackage{bm}
\usepackage{hyperref}
\usepackage[mathlines]{lineno}

\newcommand\aref[1]{Appendix~\ref{#1}}

\newcommand\eref[1]{Eq.~\ref{#1}}
\newcommand\sref[1]{Section~\ref{#1}}
\newcommand\fref[1]{Fig.~\ref{#1}}

\begin{document}

\global\long\def\ket#1{\left| #1\right\rangle }%

\global\long\def\bra#1{\left\langle #1 \right|}%

\global\long\def\kket#1{\left\Vert #1\right\rangle }%

\global\long\def\bbra#1{\left\langle #1\right\Vert }%

\global\long\def\braket#1#2{\left\langle #1\right. \left| #2 \right\rangle }%

\global\long\def\bbrakket#1#2{\left\langle #1\right. \left\Vert #2\right\rangle }%

\global\long\def\av#1{\left\langle #1 \right\rangle }%

\global\long\def\tr{\text{tr}}%

\global\long\def\Tr{\text{Tr}}%

\global\long\def\pd{\partial}%

\global\long\def\im{\text{Im}}%

\global\long\def\re{\text{Re}}%

\global\long\def\sgn{\text{sgn}}%

\global\long\def\Det{\text{Det}}%

\global\long\def\abs#1{\left|#1\right|}%

\global\long\def\up{\uparrow}%

\global\long\def\down{\downarrow}%

\global\long\def\vc#1{\mathbf{#1}}%

\global\long\def\bs#1{\boldsymbol{#1}}%

\global\long\def\t#1{\text{#1}}%

\global\long\def\ii{\mathrm{i}}%

\preprint{APS/123-QED}

\title{Mean-field dynamics of an infinite-range interacting quantum system:\\ chaos, dynamical phase transition, and localisation}

\author{Bojan \v{Z}unkovi\v{c}}
\email{bojan.zunkovic@fri.uni-lj.si}
\author{Antonio Zegarra}%
\affiliation{%
University of Ljubljana,\\ 
Faculty of Computer and Information Science,\\ 
Ljubljana, Slovenia
}%




\date{\today}

\begin{abstract}
We investigate the dynamical properties of the XY spin 1/2 chain with infinite-range transverse interactions and find a dynamical phase transition with a chaotic dynamical phase. In the latter, we find non-vanishing finite-time Lyapunov exponents and intermittent behavior signaled by fast and slow entropy growth periods. Further, we study the XY chain with a local self-consistent transverse field and observe a localization phase transition. We show that localization stabilizes the chaotic dynamical phase.
\end{abstract}

\maketitle


\section{\label{sec:level1} Introduction}
Long-range many-body quantum models are relevant for the description of the atomic, molecular, optical~\cite{blatt2012quantum,britton2012engineered,richerme2014non,baumann2010dicke} and cold atom experiments~\cite{endres2016atom,barredo2018synthetic}, with dominant long-range interactions. Besides, the tunable interaction range enables studying the transition between long- and short-range interactions~\cite{labuhn2016tunable} with drastically different behavior, e.g., regarding thermalization and information-spreading properties. We attribute the absence of thermalization in long-range interacting systems to persistent oscillations due to a mean-field-like behavior of collective degrees of freedom~\cite{ueda2020quantum}. In contrast, short-range interacting systems typically avoid thermalization only in the Anderson or many-body localized regimes~\cite{anderson1958absence, abanin2019colloquium}. Likewise, we generally bound information spreading in short-range systems by linear light cones, which we can circumvent in systems with long-range interactions leading to super-luminal information spreading~\cite{richerme2014non,mottl2012roton,jurcevic2014quasiparticle,hauke2013spread,pappalardi2018scrambling}. Long-range interacting systems also display many unique phenomena, e.g., time-translation symmetry breaking~\cite{sacha2017time,zhang2017observation,choi2017observation,rovny2018observation} and dynamical phase transition~\cite{sciolla2013quantum,piccitto2019dynamical,halimeh2017dynamical,vzunkovivc2018dynamical,halimeh2017prethermalization,vzunkovivc2016dynamical,defenu2023out}.

The latter comes in two flavours~\cite{zvyagin2016dynamical,heyl2018dynamical,defenu2023out}. First dynamical phase transition considers a dynamical order parameter, which is the time average of the standard equilibrium order parameter during a quantum quench~\cite{yuzbashyan2006relaxation,sciolla2010quantum}. We observe a transition from a vanishing to a finite dynamical order parameter upon changing the final quench parameters. The second type of a dynamical transition considers the rate function of the Loschmidt echo~\cite{heyl2013dynamical}, which displays kinks in the time evolution that can, by analogy to the standard partition function, be regarded as signatures of a (dynamical) phase transition. In some specific models, a link between the two types of dynamical phase transitions has been established~\cite{vzunkovivc2016dynamical, Zunkovic:2018aa, lerose2019impact}.  

Further, by combining long- and short-range interactions, we can engineer a \textit{chaotic} dynamical phase~\cite{lerose2018chaotic,lerose2019impact}, which we characterize by hypersensitivity of the dynamical order parameter to initial conditions and quench parameters. This phase occurs due to an unstable semiclassical paramagnetic solution and quantum correlation-induced tunneling between stable ferromagnetic solutions~\cite{lerose2019impact}. Although the fractal nature of the phase diagram in the \textit{chaotic} phase has been established~\cite {piccitto2019dynamical}, its characterization by the dynamical properties of single trajectories is still lacking. Recent work in this direction \cite{lewis2021characterizing} considers chaotic properties of the Dicke model in various dynamical phases. We complement this work by considering a different model with extended quantum degrees of freedom and studying the interplay between classical chaos, entanglement growth, and localization.

We study the XY spin 1/2 chain with infinite-range transverse interactions and establish the \textit{chaotic} nature of dynamics in the chaotic dynamical phase by numerically calculating the finite-time Lyapunov exponents in the mean-field picture.

Further, we extend the model by considering many parallel XY chains coupled by infinite-range transverse interactions. The mean-field approach leads to an XY model with a local self-consistent disordered transverse field. At each fixed time, the model is Anderson localized, which enables studying the effect of localization on the dynamical phase transition and the chaotic dynamical phase. 

In \sref{sec: global mf}, we discuss the dynamical phase transition in the XY model with a global self-consistent transverse field. In particular, we numerically calculate finite-time Lyapunov exponents in all regimes. In \sref{sec: local mf}, we discuss the XY model with a local self-consistent transverse field and show that localization stabilizes the chaotic dynamical phase. We conclude in \sref{sec: conclusions}.

\section{Global mean-field interactions}\label{sec: global mf}
In this section, we analyze the quench dynamics of an XY spin-1/2 chain with infinite-range, transverse interactions 
\begin{align}
H & =-J\sum_{i=1}^{L-1}\left(\cos\eta\,\sigma_{i}^{x}\sigma_{i+1}^{x}+\sin\eta\,\sigma_{i}^{y}\sigma_{i+1}^{y}\right)\nonumber \\
 & -\frac{g}{2L}\sum_{i,j=1}^{L}\sigma_{i}^{z}\sigma_{j}^{z}.
 \label{eq: Hamiltonian}
\end{align}
We use open boundary conditions and denote the system size by $L$, the transverse coupling by $g$, and the XY-coupling by $J$
(set to one in the following), and parametrize the anizotropy by $\eta\in\left[0,2\pi\right[$. With $\sigma^{\rm x,y,z}_j$, we represent standard Pauli matrices acting non-trivially only on-site $j$.

The model has two exactly solvable limits. Without transverse interactions, i.e., $g=0$, we obtain the XY spin-1/2 chain solvable by the Jordan-Wigner transformation to free fermions~\cite{lieb1961two}. The XY chain exhibits a quantum phase transition at $\eta=\pm \pi/4$ between two ordered phases with a non-vanishing magnetization in the $x$ and $y$ directions as the order parameter. However, the system has no finite temperature order. We describe its long-time dynamics by the generalized Gibbs ensemble~\cite{essler2016quench}. 

In the case of large transverse infinite-range coupling $g/J\gg1$, the second term in \eref{eq: Hamiltonian} dominates. At $J=0$ the model reduces to the zero-field completely anisotropic Lipkin-Meshkov-Glick (LMG) model~\cite{lipkin1965validity}. We solve the LMG model in the large-$L$ limit by the mean-field approach and find a second-order finite temperature phase transition with the order parameter $\phi^{\rm z}=\frac{1}{L}\sum_{i=1}^{L}\av{\sigma_{i}^{z}}$~\cite{botet1983large}. Despite being very simple, it has non-trivial entanglement dynamics described by the semiclassical approach~\cite{vidal2004entanglement}.

In the thermodynamic limit, $L\to\infty$, the mean-field approach also enables efficient numerical and perturbative analytical treatment of the model at intermediate values of the couplings $J$ and $g$. In this limit, the model has an intriguing phase diagram with reentrant and non-algebraic phase transitions~\cite{zunkovic23a}. Interestingly, the mean-field approach also provides insights into spectral features and eigenstates in the middle of the spectrum~\cite{den1976systems,granet2023exact,zunkovic23a}.

\subsection{Mean-field equation of motion\label{sec:meanfield}}
We follow~\cite{zunkovic23a} and describe the quench dynamics of the model in the mean-field approximation. We apply the decoupling ansatz $\sigma_{i}^{z}\sigma_{i}^{z}\to\av{\sigma_{i}^{z}}\sigma_{i}^{z}+\sigma_{i}^{z}\av{\sigma_{i}^{z}}-\av{\sigma_{i}^{z}}\av{\sigma_{i}^{z}}$ and reduce the 
Hamiltonian in \eref{eq: Hamiltonian} to an XY chain in a self-consistent transverse field 
\begin{align}
H_{\text{MF}}= & -J\sum_{i=1}^{L-1}\left(\cos\eta\,\sigma_{i}^{x}\sigma_{i+1}^{x}+\sin\eta\,\sigma_{i}^{y}\sigma_{i+1}^{y}\right)\nonumber \\
 & -h\sum_{i=1}^{L}\sigma_{i}^{z}+\frac{1}{2g}h^{2},\label{eq: mean-field H}
\end{align}
with the self-consistency condition
\begin{align}
h= & \frac{g}{L}\sum_{i=1}^{L}\av{\sigma_{i}^{z}}.\label{eq:selfcond}
\end{align}
Fixing $h$, the mean-field Hamiltonian $H_{\text{MF}}$ admits an exact solution by Jordan-Wigner transformation implemented by  $a_{r}^{\dagger}=e^{\ii\pi\sum_{r'=0}^{r-1}\sigma_{r'}^{+}\sigma_{r'}^{-}}\sigma_{r}^{+}$. The Jordan-Wigner fermions $a_{r}^{(\dag)}$ obey fermionic commutation relations, $\left[a_{r},a_{r'}^{\dagger}\right]=\delta_{rr'}$. We solve the fermionic Hamiltonian in the momentum basis $a_{q}^{\dagger}=\frac{1}{\sqrt{L}}\sum_{r}e^{-\ii qr}a_{r}^{\dagger}$. The mean-field Hamiltonian $H_{\t{MF}}$ and the self-consistency condition in the momentum basis simplify to 
\begin{align}
\label{eq: pairing H}
H_{\t{MF}}(t)& =H_{\t{XY}}-h(t)\frac{1}{L}\sum_{q}\left(2a_{q}^{\dagger}a_{q}-1\right)+\frac{L}{2g}h(t)^{2},\\
\label{eq: pairing h}
h(t)&=\frac{g}{L}\sum_q\left\langle2a^\dag_qa_q-1\right\rangle,
\end{align}
where
\begin{align}
    H_{\t{XY}}=-\sum_q &(cs_+ \cos q +h)\left(a_q^\dag a_q+a_{-q}^\dag a_{-q}\right) \\ \nonumber -{\rm i}\sum_q &cs_-\sin q \left(a^\dag_{-q}a^\dag_{q}+a_{-q}a_q\right)+ L h
\end{align}
is the Fourier transformed XY model and $cs_-=J(\cos\eta-\sin\eta),\quad cs_+=J(\cos\eta+\sin\eta)$.

Finally, we write the Hamiltonian (\eref{eq: pairing H}) and the self-consistency condition (\eref{eq: pairing h}) in terms of Anderson\textquoteright s
pseudospin representation $\tau^{\alpha}_{q}=\psi_{q}^{\dagger}\tau^{\alpha}\psi_{q}$. The Nambu spinor is $\psi_{q}=\left( a_{q},a_{-q}^{\dagger}\right) ^{T}$ and $\tau^{\alpha}$ denotes standard Pauli matrices with $\alpha=\t{x,y,z}$. The pseudospin satisfies the commutation relations for angular momentum, $\left[\tau_{q}^{\alpha},\tau_{q'}^{\beta}\right]=2i\delta_{qq'}\varepsilon_{\alpha\beta\gamma}\tau_{q}^{\gamma}$.
In the pseudo-spin representation the mean-field Hamiltonian (\eref{eq: mean-field H}) simplifies to
\begin{align}
H_{\t{MF}}\left(t\right)=\sum_{q}\vec{B}_{q}\left(t\right)\cdot\vec{\tau}_{q}\label{eq: Hamiltonian pseudospin},
\end{align}
with a time-dependent magnetic field
\begin{align}
\vec{B}_{q}\left(t\right) & =\left\{ 0,4\sin(q)cs_-,4\left[\cos(q)cs_++h(t)\right]\right\} ^{T},
\end{align}
the self-consistency condition $ h(t)=g \phi^{\rm z}(t)$, and the time dependent order parameter $\phi^{\rm z}(t)=\sum_q\av{\tau^{\rm z}_q(t)}/L$.

Since the pseudo-spin Hamiltonian is linear, the self-consistent dynamics of the pseudo-magnetization vector $\vec{\phi}_{q}\left(t\right)=\bra{\phi\left(t\right)}\vec{\tau}_{q}\ket{\phi\left(t\right)}$ simplifies to
\begin{align}
\pd_{t}\vec{\phi}_{q}\left(t\right) & =\ii\bra{\vec{\phi}\left(t\right)}\left[H_{\t{MF}}\left(t\right),\vec{\tau}_{q}\right]\ket{\vec{\phi}\left(t\right)}\nonumber \\
 & =\vec{B}_{q}\left(t\right)\times\vec{\phi}_{q}\left(t\right).\label{eq:phi_of_t}
\end{align}
The simplicity of the final non-linear mean-field dynamical equations (\eref{eq:phi_of_t}) enables us to numerically and analytically study the quench dynamics of the time-dependent order parameter,  the entanglement entropy, and the Lyapunov exponents.

\subsection{Stability and Lyapunov exponents}
We aim to find a correspondence between the dynamical phase transition and non-vanishing Lyapunov exponents in the mean-field picture. We determine the finite-time Lyapunov exponents by adopting the Bennetin algorithm~\cite{benettin1980lyapunov} (see also \aref{app:benettin}), which evolves a set of displaced trajectories with linearized equations of motion determined by the Jacobian. The latter is for a specific configuration $\{\vec{\phi}_{q_i}\}$, with $i=1,\ldots L/2$ and $q_i = \frac{2\pi i}{L}$, given by
\begin{align}
    \nonumber
    J&=\begin{bmatrix}
        A_{1}+B_1       & B_1 & B_1 & \dots & B_1 \\
        B_2       & A_2+B_2 & B_2 & \dots & B_2 \\
        \hdotsfor{5} \\ 
        B_{L/2}       & B_{L/2} & B_{L/2} & \dots & A_{L/2}+B_{L/2}
    \end{bmatrix},\\ \nonumber
    A_i&=\begin{bmatrix}
        0 & 4(cs_+\cos q_i+h) & -4cs_-\sin q_i \\
        -4(cs_+\cos q_i+h) & 0&0 \\
        4cs_-\sin q_i &0 &0 \\
    \end{bmatrix},\\ 
    B_i&=\begin{bmatrix}
        0 & 0 & -\frac{4g}{L} \phi_{q_i}^{ \rm y} \\
        0 & 0 & \frac{4g}{L} \phi_{q_i}^{ \rm x} \\
        0 & 0 &0 \\
    \end{bmatrix}.
\end{align}

First, we focus on the ground state fixed point $\vec{\phi}^{\rm fixed}_q=-\vec{B}_q/\lVert \vec{B}_q \rVert_2$ with $h=0$, which satisfies the self-consistency condition. The Jacobian eigenvalues determine the stability of the fixed point. In the case where we quench only the infinite-range interaction strength $g$ and keep the interaction angle fixed at zero, i.e., $\eta_{\rm initial}=\eta_{\rm final}=0$, we find the exact spectrum of the Jacobian. Assuming $L$ is even and $L\geq4$, we find $L-1$ imaginary eigenvalues $\pm 4\ii$, $(L-2)/2$  eigenvalues zero, and one pair of eigenvalues $\lambda_{\pm}=\pm2\sqrt{2}\sqrt{g-2}$. The Jacobian has a similar spectrum for all quenches to the Ising model, i.e. $\eta_{\rm initial}\neq\eta_{\rm final}=0$. We numerically find $L-1$ imaginary eigenvalues $\pm 4\ii$, $(L-2)/2$ eigenvalues zero, and two eigenvalues $\pm 2\sqrt{2}\sqrt{g-g^*}$, where $g^*$ is the numerically determined dynamical critical point. For quenches with a fixed interaction angle $\eta_{\rm initial}=\eta_{\rm final}=\eta$ the dynamical critical point is determined by 
\begin{align}
    g^*/J & =\frac{\pi\sin(2\eta)}{\sqrt{1-\sin(2\eta)}\left[E(\tilde{\eta})-K(\tilde{\eta})\right]},\label{eq: critical g}\\
    \tilde{\eta} & =2+\frac{2}{\sin(2\eta)-1}\nonumber,
\end{align}
where $E(x)$ and $K(x)$ are the complete elliptic integrals of the first and the second kind. The dynamical criticality point corresponds to the onset of instability of the paramagnetic solution to the free energy~\cite{zunkovic23a}.

However, the stability of the initial fixed point does not determine the long-time dynamical properties of neighboring trajectories, i.e., Lyapunov exponents. In particular, the long-time regular motion of the order parameter in the dynamical ferromagnetic phase suggests vanishing Lyapunov exponents. On the other hand, the \textit{chaotic} dynamical phase~\cite{lerose2018chaotic} displays a hypersensitivity of the dynamical order parameter to the initial conditions, leading to the fractalization of the phase diagram~\cite{strzalko2008dynamics,piccitto2019dynamical} indicating non-vanishing Lyapunov exponents. However, the calculation of the Lyapunov exponents in this region has been out of reach due to the extensive computational effort necessary to determine even the dynamics of models displaying the chaotic dynamical phase. An exception is a recent study of the Dicke model, where we couple one bosonic mode to a big spin~\cite{lewis2021characterizing}. In the following, we will utilize the simplicity of the dynamical equations and study the finite-time Lyapunov exponents in all dynamical phases. 

\subsection{Dynamical phase transition \label{sec:dpt}}
In this section, we will study the self-consistent, non-linear, quench dynamics of the pseudo-magnetization vector $\vec{\phi}_q(t)$ and the time-dependent order parameter $\phi^{\rm z}(t)$. In quench dynamics, we start in a ground state of a Hamiltonian with an initial set of parameters and then evolve the state with another set of parameters. We instantly switch from the initial to the final set of parameters and keep the latter fixed during the evolution. We will quench the infinite-range transverse interaction strength $g$ and the interaction angle $\eta$. The initial state will always have $g=0$ and correspond to a paramagnetic state with a vanishing time-dependent order parameter $\phi^{\t z}(0)$. However, since the paramagnetic ground state with a vanishing time-dependent order parameter is a fixed point of the dynamics for any $g$, we start with a slightly perturbed ground state. Our initial state is the ground state of an XY chain with a small transverse field $h=\epsilon$. Unless specified, we set $\epsilon=10^{-4}$.

\begin{figure}[!htb]
    \centering
    \includegraphics[width=\columnwidth]{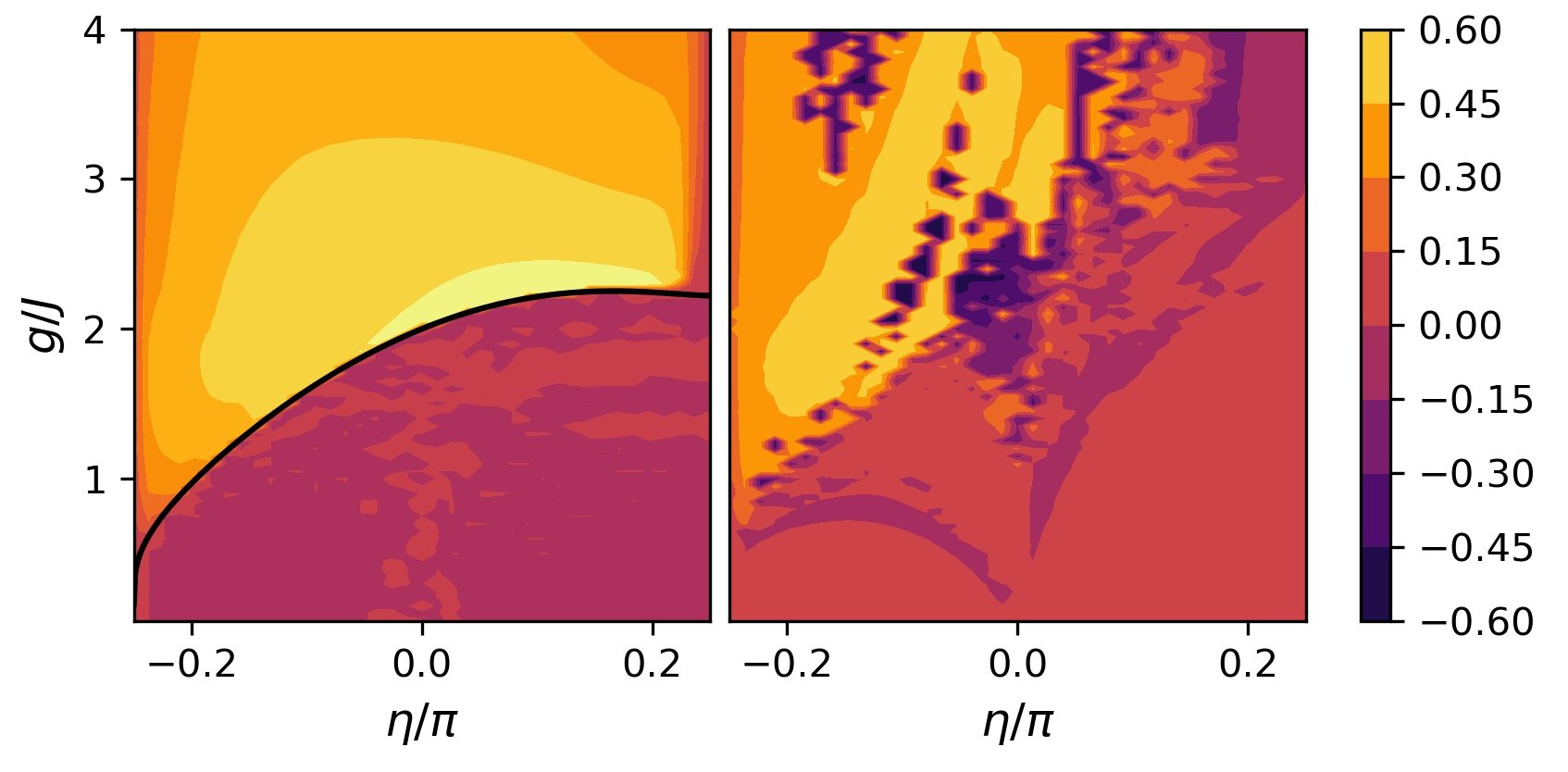}
    \caption{We show the dynamical order parameter $\overline{\phi^{\rm z}}$ after a quantum quench. Left panel corresponds to the quench $(\eta,0)\rightarrow(\eta,g)$ and the right panel corresponds to the quench $(-\pi/4,0)\rightarrow(\eta,g)$. Interestingly, we do not observe a chaotic dynamical order for a quench with a constant interaction angle $\eta$. The black line in the left panel shows the analytical result for the critical transverse interaction $g^*$ determined by \eref{eq: critical g}. Simulation parameters: $L=1000$, $T=100$.}
    \label{fig:2d phase diagram}
\end{figure}

In \fref{fig:2d phase diagram} we show a phase diagram of the dynamical order parameter after a quantum quench, namely $\overline{\phi^{\rm z}}=\frac{1}{T}\int_{t=0}^T\phi^{\rm z}(t)$, where $T$ is the total simulation time which should in principle go to infinity. The left panel corresponds to a quantum quench with a constant interaction angle $\eta_{\rm initial}=\eta_{\rm final}=\eta$ and the right panel to a quench with $\eta_{\rm initial}=-\pi/4,~\eta_{\rm final}=\eta$. In the first case, we find a standard dynamical phase transition from the dynamical paramagnetic to the dynamical ferromagnetic phase with the critical field $g^*$ determined by \eref{eq: critical g}. In the second case, we also observe a chaotic dynamical phase, where the time-dependent order parameter becomes highly sensitive to the final quench parameters.

\begin{figure}[!htb]
    \centering
    \includegraphics[width=0.8\columnwidth]{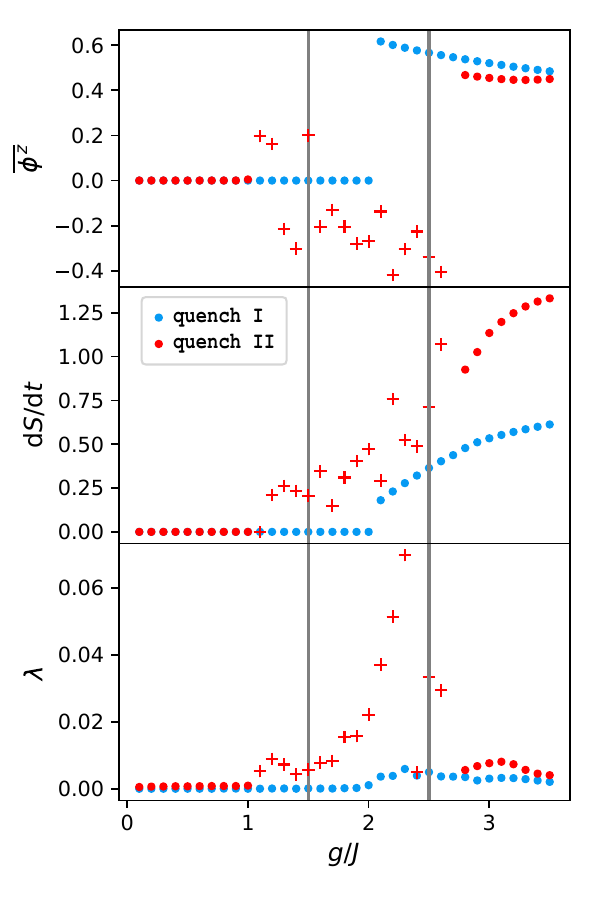}
    \caption{We show a dynamical phase transition in two quench scenarios -- \texttt{quench I}: $(0,0)\rightarrow (0,g)$ and \texttt{quench II}: $(-\pi/4,0)\rightarrow (0,g)$. The top plot shows the order parameter $\overline{\phi^{\t z}}$, the middle plot shows the long-time entropy growth with time, and the bottom plot shows the largest finite-time Lyapunov exponent at the final simulation time $T$. In the \texttt{quench I} (blue points), we observe a dynamical phase transition at the critical $g=2$. The order parameter and entropy display a first-order phase transition, while the finite-time Lyapunov exponent remains zero in both phases. We nonetheless obtain a small finite-time Lyapunov exponent in the ferromagnetic case due to a slower convergence with simulation time $T$(see also \fref{fig:dpt trajectories} and \aref{app:dpt}). In the \texttt{quench II}, we observe a transition from a paramagnetic dynamical phase at small $g$ to a chaotic dynamical phase. The chaotic phase is revealed by a high sensitivity of the order parameter and entropy growth with $g$. Accordingly, the finite-time Lyapunov exponent converges to a \textit{non-vanishing} value in the \textit{chaotic} dynamical phase. With plus markers, we denote results in the chaotic dynamical region, where the convergence of the order parameter and entropy growth with system size is hampered by the high sensitivity to the initial conditions. Due to computational complexity, we calculated the finite-time Lyapunov exponents for a smaller system size $L=300$ and longer time $T=10^4$. We obtain the order parameter and the entropy data from simulations with $L=1000$ and $T=100$.}
    \label{fig:dpt slices}
\end{figure}

\begin{figure}[!htb]
    \centering
    \includegraphics[width=0.95\columnwidth]{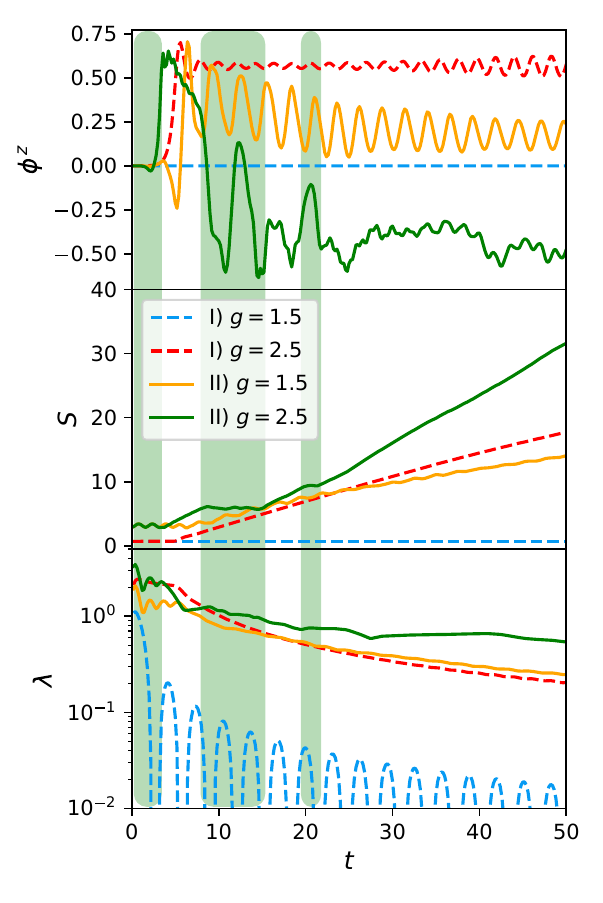}
    \caption{Typical trajectories of the order-parameter $\phi^{\rm z}(t)$, entanglement entropy $S(t)$ and the finite-time Lyapunov exponent $\lambda(t)$. We show the quenches \texttt{quench I} and \texttt{quench I}, denoted in the legend by I) and II), respectively. The quench in the paramagnetic phase ( \texttt{quench I} $g=1.5$) displays regular motion with constant entropy. The quench to the ferromagnetic phase (\texttt{quench I} $g=2.5$)  has a finite order parameter with regular motion and linear growing entropy. The \texttt{quench II} $g=2.5$ (green lines) to the chaotic dynamical phase displays intermittent behavior, switching between regular motion with linear entropy growth and irregular motion with constant entropy. We mark the irregular regimes by the three green-shaded regions. In these regimes, the order parameter changes drastically (even a sign) and the entropy remains constant. The effect in the finite-time Lyapunov exponent is less pronounced and delayed due to time averaging. In the regular regimes, the order parameter experiences small oscillations and the entropy grows linearly with time. In the \texttt{quench II} $g=1.5$, the intermittent behavior is absent, we observe regular oscillations in the order parameter and linear growth of the entropy. In both cases, the finite-time Lyapunov exponent seems to converge to a non-vanishing value. All quantities have been calculated with the system size $L=1000$.}
    \label{fig:dpt trajectories}
\end{figure}

In the following, we focus on quenches to the Ising model, i.e., $\eta_{\rm final}=0$,
\begin{itemize}
\item \texttt{quench I}: $(0,0)\rightarrow (0,g)$,
\item \texttt{quench II}: $(-\pi/4,0)\rightarrow (0,g)$,
\end{itemize}
and study the dynamics of the time-dependent order parameter $\phi^{\rm z}$, the half-chain entanglement entropy $S(t)$, and the largest finite-time Lyapunov exponent $\lambda(t)$ as obtained from the Bennetin algorithm (see \fref{fig:dpt slices}). We observe that the convergence of the order parameter in the chaotic dynamical phase requires a large system size for which we can not calculate the finite-time Lyapunov exponents. The latter is, therefore, calculated with a smaller system size. More details on the convergence of the results are discussed in \aref{app:dpt}. In the \texttt{quench I}, we observe a first-order dynamical phase transition in $\overline{\phi^{\rm z}}$, which is accompanied by a first-order transition in the long-time entropy growth ${\rm d}S/{\rm d}t$. Below the critical point $g^*=2$, the dynamics are essentially described by a product state whose transverse magnetization vanishes. Above the critical point in the dynamical ferromagnetic phase, the long-time entropy growth increases with $g$, and the dynamical order parameter decreases with $g$ (see also \fref{fig:dpt trajectories}). In the \texttt{quench I}, the largest finite-time Lyapunov exponent vanishes, which is consistent with the observed regularity of the trajectories of $\phi^{\rm z}(t)$ (see also \fref{fig:dpt trajectories}). 

In the \texttt{quench II}, we observe a transition to the chaotic dynamical phase with high sensitivity of the dynamical order parameter and the long-time entropy growth with respect the the final quench parameters. We numerically find \textit{non-vanishing} finite-time Lyapunov exponents in the \textit{chaotic} dynamical phase.

In \fref{fig:dpt trajectories}, we show typical behavior of the time-dependent order parameter, the entanglement entropy, and the finite-time Lyapunov exponent in different regimes. The dynamical paramagnetic trajectory has vanishing order parameter and entropy and $1/t$ convergence of the finite-time Lyapunov exponent typical for a regular motion. In the ferromagnetic case, the order parameter initially increases exponentially with the instability exponent of the fixed point $\lambda_{\rm fixed}$ up to time $\sim \log \epsilon/\lambda_{\rm fixed}$. In this regime, the entropy remains constant, and the finite-time Lyapunov exponent converges to $\lambda_{\rm fixed}$. After the initial exponential growth, the dynamical order oscillates around its final value, the entanglement entropy grows linearly with time, and the numerical approximation to the finite-time Lyapunov exponent decreases as $1/t$, consistent with regular motion.

Interestingly, in the \textit{chaotic} dynamical phase, the system displays a dichotomy of classical and quantum motion (see \fref{fig:dpt trajectories}). First, the time-dependent order parameter $\phi^{\rm z}(t)$ increases exponentially as in the dynamical ferromagnetic phase discussed above. Similarly, the entropy is constant, and the finite-time Lyapunov exponent converges to the largest eigenvalue of the Jacobian at the fixed point. After the initial growth, the dynamics of the order parameter of some trajectories becomes intermittent, i.e., it displays sudden irregular transitions between short-time regular motion. During the regular motion, the entropy grows linearly, and the numerical approximation to the finite-time Lyapunov exponent decreases. The entropy remains constant or decreases during the transitions, whereas the finite-time Lyapunov exponent typically increases. 

Finally, many trajectories in the chaotic dynamical phase do not display intermittent behavior. Instead, the dynamical order parameter shows a bounded irregular motion accompanied by linear entropy growth and a non-vanishing finite-time Lyapunov exponent.

\subsection{Summary and discussion}
We study the dynamical phase transition in the XY model with a mean-field self-consistent transverse field. For quenches with constant interaction angle $\eta$, we calculate the exact critical transverse interaction $g^*$, which determines the transition to a standard ferromagnetic dynamical phase with linear entropy growth and vanishing order parameter. In cases with quenched interaction angle, we find a chaotic dynamical phase associated with hypersensitivity of the dynamical order parameter to the quench parameters~\cite{lerose2018chaotic}. We numerically show a non-vanishing finite-time Lyapunov exponent in the chaotic dynamical phase. The finite-time Lyapunov exponent vanishes in the paramagnetic and standard ferromagnetic phases. 

Further, for some initial conditions, we observe an intermittent classical-quantum dichotomy. Namely, the order parameter follows short periods of regular motion followed by \textit{chaotic} transitions. The entanglement entropy grows linearly, and the finite-time Lyapunov exponent drops in the regular parts of the trajectory. During transitions between the regular parts, the finite-time Lyapunov exponent typically increases, and the entanglement entropy remains constant or even drops. We contrast this phenomenon with the semiclassical theory in long-range systems, where linear entropy growth is associated with the exponential divergence of the semiclassical trajectories~\cite{lerose2020bridging,lerose2020origin}.

A similar chaotic dynamical phase has been described in the Dicke model~\cite{lewis2021characterizing}, where the chaotic dynamical phase has been linked to quantum and classical Lyapunov exponents~\cite{lewis2019unifying,alavirad2019scrambling,chavez2019quantum}. However, no intermittent behavior of entropy and order-parameter has been reported.

First experimental observations of dynamical phase transitions have been demonstrated in a trapped
ion quantum simulator consisting of a chain of up to $N = 53$ spins~\cite{zhang2017observation} that enables a realization of long-range Ising interactions. Further, 
\cite{muniz2020exploring} realized the Lipkin-Meshkov-Glick model in a  cavity-QED simulator using
ensembles of $N \approx 105 –106$ atoms and observed a dynamical phase transition in the order parameter. However, a chaotic dynamical phase requires integrability breaking terms (or the interplay between short- and long-range interactions) and has not been observed so far in these platforms. Nevertheless, detecting the signature of the chaotic dynamical phase seems feasible by realizing the Dicke model in trapped ions simulators~\cite{lewis2021characterizing}.

\section{Local mean-field interactions}\label{sec: local mf}
In this section, we study the mean-field dynamics of an array of XY chains, which are connected with an infinite-range transverse interaction
\begin{align}
H  =&-J\sum_{i=1}^{L-1}\sum_{\mu=1}^{N}\left(\cos\eta\,\sigma_{i,\mu}^{x}\sigma_{i+1,\mu}^{x}+\sin\eta\,\sigma_{i,\mu}^{y}\sigma_{i+1,\mu}^{y,\mu}\right)\nonumber \\
 & -\frac{g}{2N}\sum_{i=1}^L\sum_{\mu,\nu=1}^{N}\sigma_{i,\mu}^{z}\sigma_{i,\nu}^{z}.
 \label{eq: Hamiltonian 2d}
\end{align}
The first Latin index corresponds to the site along the chain, and the second Greek index corresponds to the index of the chain. At each site, the spin is transversely coupled with other chains at the same position. If the system is in a homogeneous state, the dynamics of the models defined by \eref{eq: Hamiltonian} and \eref{eq: Hamiltonian 2d} agree. However, if we introduce a small perturbation, the systems behave differently. While the dynamics of the model discussed in the previous section remain qualitatively unchanged by a small perturbation, the phenomenon of (Anderson) localization in the XY model drastically changes the behavior of coupled chains introduced above.

We utilize the same mean-field approach as in the previous section~\sref{sec:meanfield}. After inserting the decoupling ansatz into the model~\eref{eq: Hamiltonian 2d} we obtain the mean-field Hamiltonian
\begin{align}
\label{eq: mean-field H 2d}
H_{\rm MF}=&\sum_{\mu=1}^NH^{\mu}_{\rm MF}, \\ \nonumber
H^\mu_{\text{MF}}= & -J\sum_{i=1}^{L-1}\left(\cos\eta\,\sigma_{i}^{x}\sigma_{i+1,\mu}^{x}+\sin\eta\,\sigma_{i,\mu}^{y}\sigma_{i+1,\mu}^{y}\right)\nonumber \\ \nonumber
 & -\sum_{i=1}^{L}\left({h_i\sigma_{i,\mu}^{z}+\frac{1}{2g}h_i^{2}}\right),
\end{align}
with the \textit{local} self-consistency condition
\begin{align}
h_i= & \frac{g}{N}\sum_{\mu=1}^{N}\av{\sigma_{i,\mu}^{z}}.\label{eq: selfcond 2d}
\end{align}
The mean-field Hamiltonian \eref{eq: mean-field H 2d} describes an array of non-interacting XY spin-1/2 chains coupled through the \textit{local} self-consistency condition \eref{eq: selfcond 2d}. Therefore, we can describe the dynamics of chain-independent initial conditions by a single XY chain with the Hamiltonian
\begin{align}
    H_{\text{MF}}= & -J\sum_{i=1}^{L-1}\left(\cos\eta\,\sigma_{i}^{x}\sigma_{i+1}^{x}+\sin\eta\,\sigma_{i}^{y}\sigma_{i+1}^{y}\right)\nonumber \\ \label{eq: mean-field H local}
 & -\sum_{i=1}^{L}\left({h_i\sigma_{i}^{z}+\frac{1}{2g}h_i^{2}}\right),
\end{align}
and the \textit{local} self-consistency condition
\begin{align}
h_i= & g\av{\sigma_{i}^{z}}.\label{eq: selfcond 2d local}
\end{align}
The difference between the \textit{local} and the \textit{global} self-consistency condition is that the former introduces an effective disorder in the transverse field, which in 1D leads to Anderson localization in the thermodynamic limit for any disorder strength~\cite{anderson1958absence,thouless1972relation}.

We will study the effect of Anderson localization in 1D on the dynamical phase transition discussed in the previous section. Conversely, we will also study if the mean-field \textit{local} self-consistency condition induces the (Anderson) localization transition at finite disorder. 

\subsection{Dynamical phase transition \label{sec: local mf}}
In this section, we repeat the analysis of \sref{sec:dpt} with a slightly modified initial condition. Namely, we use a ground state of an XY Hamiltonian with a small site-dependent local transverse field $h_i\in[0,\epsilon]$ instead of a homogeneous one. As in the homogeneous case, we test the \texttt{quenches I} and \texttt{II}. Since the numerical simulations are more demanding due to the local self-consistency condition of the transverse field $h_i$, we calculate the time-dependent order parameter, the half-chain entanglement entropy, and the standard deviation of the transverse magnetization profile. Notably, we omit the calculation of the Lyapunov exponents. 

\begin{figure}[!htb]
    \centering
    \includegraphics[width=\columnwidth]{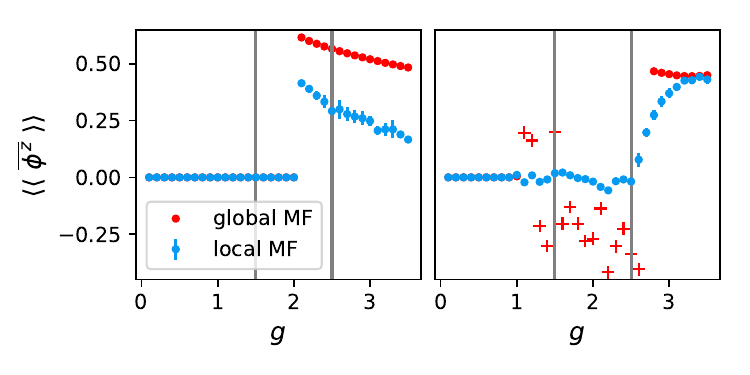}
    \caption{Dynamical phase diagram for quenches \texttt{quench I} (left) and \texttt{quench II} (right). With $\langle\langle \overline{\phi^{\rm z}}\rangle\rangle$ we denote the disorder average of the long-time average of the time-dependent order parameter $\overline{\phi^{\rm z}}$. The blue dots with error bars represent the mean and the standard deviation of the dynamical order parameter over ten realizations of the initial conditions. For comparison, we also show the homogeneous global mean-field phase diagram (red dots). The \texttt{quench I} does not change qualitatively. In contrast, in the \texttt{quench II}, we do not observe high sensitivity of the order parameter to $g_{\rm final}$ as in the homogeneous global mean-field case. Simulation parameters: $L=400$, $t=400$.}
    \label{fig:local-MF phase diagram}
\end{figure}

In \fref{fig:local-MF phase diagram}, we show the phase diagram of the dynamical order parameter averaged over ten initial disorder realizations. In the \texttt{quench I}, we observe the same dynamical phase transition in the dynamical order parameter at the Ising critical point $g_{\rm c}=2$ with a slightly reduced ferromagnetic order. In the \texttt{quench II}, the chaotic dynamical phase observed in the homogeneous global mean-field model disappears. At small $g$ we have a paramagnetic phase, which borders to a ferromagnetic phase at a critical point $g_c\approx 1$. After the critical point, the order parameter remains small and does not abruptly change as in the homogeneous (global mean-field) case. Finally, at $g>2.5$, the order parameter increases to its strong-coupling value $\sim 0.4$. 

\begin{figure}[!htb]
    \centering
    \includegraphics[width=0.9\columnwidth]{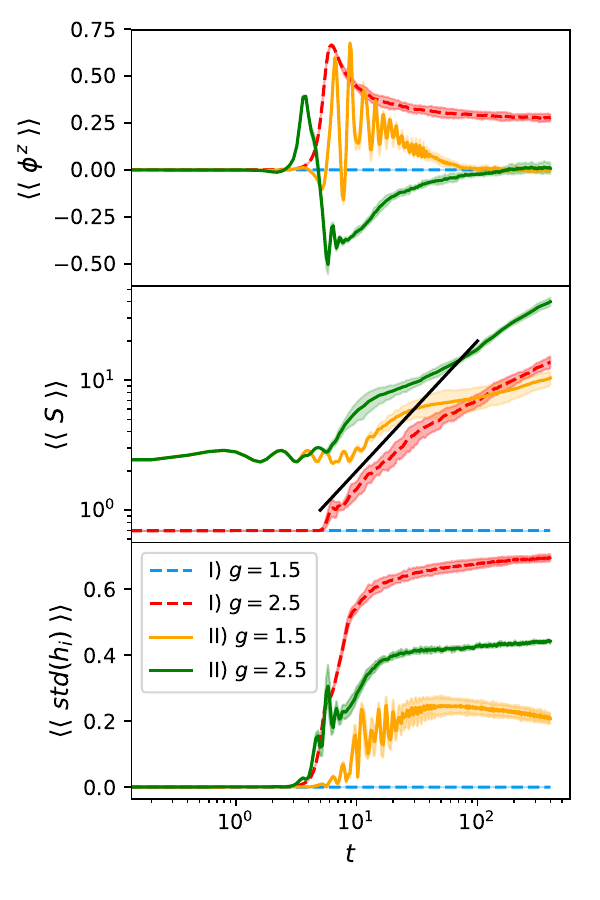}
    \caption{We show the time dependence of the mean of the order-parameter $\phi^{\rm z}(t)$ (top), entanglement entropy $S(t)$ (middle), and the standard deviation of the transverse magnetization (bottom) over ten initializations. The shaded regions denote the standard deviation of the quantities over the initializations. The qualitative behavior of the order parameter is the same for both quenches, i.e., \texttt{quench I} and \texttt{quench II}. In contrast to the global mean field, we do not observe long-time oscillations. The order-parameter trajectories converge to the dynamical order parameter and do not drastically depend on the initial state. The localization stabilizes the dynamically chaotic phase. The entropy growth slows down at longer times and is sublinear -- the black line shows linear growth for convenience. The intra-site variance of the transverse magnetization changes only slowly after the initial growth. Simulation parameters: $L=400$.}
    \label{fig:local-MF time dependence}
\end{figure}

We better understand the stabilization of the chaotic dynamical phase by observing the dynamics of relevant quantities. In \fref{fig:local-MF time dependence}, we show the time dependence of the order parameter, the entanglement entropy, and the standard deviation of the transverse magnetization profile. Both quenches (\texttt{I} and \texttt{II}) display similar behavior. In the paramagnetic phase, all quantities remain close to zero. In the ferromagnetic case, we determine two stages of the dynamics. First, in the chaotic stage, we observe an exponential increase in the order parameter due to the instability of the initial condition, where the entropy remains constant and the standard deviation remains small. After the initial short-time growth, the order parameter and the magnetization profile standard deviation peak, while the entropy increases. In the second (localized) stage, we observe a slow relaxation toward the final dynamical order parameter and a slower sublinear entropy growth. In contrast, the standard deviation of the transverse magnetization remains close to its short-time peak value. 

We observe two differences with the homogeneous case. First, the dynamics of the time-dependent order parameter in the previously chaotic region is regular, as the final order parameter converges to the long-time limit with diminishing oscillations. Also, different initializations lead to very similar dynamics of the order parameter, which is quantified by a small standard deviation at long times shown as shaded regions in \fref{fig:local-MF time dependence}.  

The second main difference is the sublinear long-time entropy growth. Unfortunately, our simulation times are insufficient to exactly determine the long-time entropy behavior. Nevertheless, our simulations suggest that the entropy likely increases as a power law $S(t)\propto t^\alpha$ with $\alpha<1$. Further convergence results of the long-time entropy growth are shown in \aref{app: local mf}.

\subsection{Domain wall initial condition}
Finally, we study spin transport properties in the XX model ($\eta=\pi/4$) with a local self-consistency condition. We induce spin transport by starting from an inhomogeneous domain-wall initial condition (strong non-equilibrium), where the left half of the chain is polarised in the positive $z$ direction and the right half of the chain in the negative $z$ direction. In the case $g=0$, we find ballistic transport consistent with the integrability of the model~\cite{lieb1961two}. We expect that above some critical $g^*$, the disorder induced by infinite-range interactions leads to (Anderson) localization and prohibits transport. 

\begin{figure}[!!htb]
    \centering
    \includegraphics[width=\columnwidth]{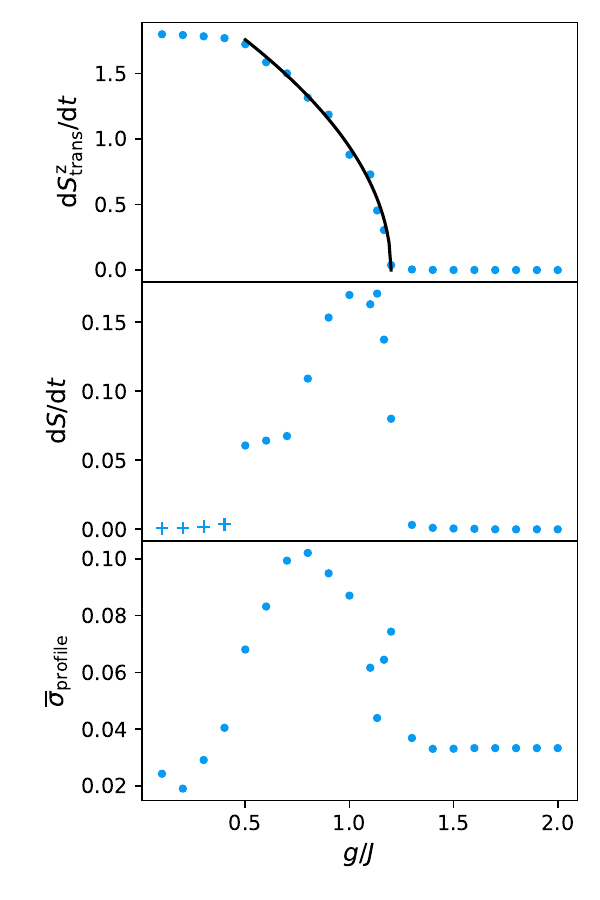}
    \caption{We show the long-time current in the middle of the chain (top), the long-time entropy growth (middle), and the long-time average of the standard deviation of the magnetization around the linear domain wall profile (bottom). The numerically determined critical point is $g^*\approx 1.2$, and the critical exponent (shown with the black line) of the long-time current is consistent with the mean-field value $1/2$. Considering the entropy growth, we find at small $g$ a logarithmic increase consistent with the free XX model. After the perturbative critical time proportional to $1/g^2$, we observe a transition to a linear long-time increase. The plus markers in the second plot indicate that the simulation times are below that critical time. While the entropy growth is still not converged, the magnetization current and the linear magnetization profile are. Therefore, we do not use the plus markers in those plots. Simulation parameters: $L=2000$, $t=340$.}
    \label{fig:dw phase diagram}
\end{figure}

In \fref{fig:dw phase diagram}, we show the long-time magnetization current ${\rm d}S^{\rm z}_{\rm trans}/{\rm d}t$, the entropy growth, and the time-averaged standard deviation $\bar{\sigma}_{\rm profile}$ of the difference between the actual domain wall magnetization and a fitted linear domain wall profile. We observe that the transferred magnetization $S^{\rm z}_{\rm trans}$ and the entropy increase linearly with time until a critical $g^*\approx 1.2$ above which they remain constant. The spin current vanishes with the standard mean-field critical exponent $1/2$. Interestingly, the entropy growth seems to slow down close to the transition. The domain wall profile is approximately linear (see also \aref{app:dw}), which can be quantified by observing the standard deviation $\sigma_{\rm profile}$ of the local magnetization around a fitted linear domain wall profile which remains small at all times.

In contrast to the homogeneous quench discussed in \sref{sec: local mf}, we find a time-independent entropy above the localization transition, which is consistent with the Anderson localized phase.

\section{Conclusions}\label{sec: conclusions}
We studied the quench dynamics of the XY model with infinite-range transverse interactions. We apply the mean-field approach, where the problem becomes analytically and numerically tractable with standard free-fermion techniques. We analytically calculate the critical point for the dynamical phase transition. Numerically, we show the presence of a \textit{chaotic} dynamical phase, which has until now been quantified only with the fractal structure of the phase diagram. We add the calculation of finite-time Lyapunov exponents to this description and show that they remain \textit{finite} only in the \textit{chaotic} dynamical phase. Our numerics suggest non-vanishing finite-time Lyapunov exponents also in the thermodynamic limit. In the \textit{chaotic} dynamical phase, we observe intermittent behavior reflected in the order parameter, entropy, and the finite-time Lyapunov exponents. During the regular motion of the order parameter, the entropy growth is linear, and the finite-time Lyapunov exponent decreases. During the chaotic order parameter motion, the entropy remains constant and the finite-time Lyapunov exponent typically increases. We contrast this connection to the linear entropy growth induced by local instabilities of the classical trajectories in long-range spin models \cite{lerose2020origin}. 

We also studied the effect of a local self-consistency condition on the dynamics of the model. At a fixed time, the mean-field approach generally leads to an XY model with a disordered (self-consistent) magnetic field, which induces (Anderson) localization. We still observe a dynamical phase transition with the same critical point. However, the chaotic dynamical phase is stabilized by localization. The time-dependent order parameter is constant at long times, and its time average does not display high sensitivity to the quench parameters.  

Finally, analyzing an inhomogeneous quench from a domain wall initial condition in the XX model, we observe a ballistic to insulator (localization) phase transition with the critical point $g^*_{\rm loc}\approx1.2$ which is smaller than the dynamical critical point $g^*_{\rm dyn}\approx 2.2$. This corroborates the absence of the chaotic dynamical phase observed in the homogeneous quench.  

In further work, it would be interesting to study the effect of quantum correlations on the observed phenomena. We can include first-order quantum corrections by the time-dependent Holstein-Primakoff approach to long-range interacting quantum systems \cite{lerose2018chaotic, piccitto2019dynamical}.

\begin{acknowledgments}
This work has been supported by the Slovenian Research Agency (ARRS) project J1-2480. Computational resources were provided by SLING – a Slovenian national supercomputing network.
\end{acknowledgments}

\bibliography{xy-zz_dynamics}

\onecolumngrid
\appendix
\section{Benettin algorithm and finite-time Lyapunov exponents \label{app:benettin}}
We apply the Benetting algorithm to calculate the Lyapunov exponents. Here, we provide a summary of the algorithm. Consider the dynamical equation
\begin{align}
    \partial_t x_i(t) = f_i(x(t)),
\end{align}
where $x$ is a vector and $f_i(x)$ are differentiable functions determining the evolution of the $i$-th vector component. Our mean-field evolution equation \eref{eq:phi_of_t} in the main text is of that form. The Lyapunov spectrum can then be determined by studying the evolution of tangent/displacement vectors $Y$, which is determined by the Jacobian
\begin{align}
    \partial_t Y(t) = J(t) Y(t), \quad J_{i,j}(t) = \frac{\partial f_i(x)}{\partial x_j}\Bigg|_{x=x(t)}.
    \label{eq:linearised lyap}
\end{align}
The initial matrix $Y(t=0)$ is an $N\times M$ matrix of $M$ random orthonormal vectors of size $N$, where $N$ is the number of degrees of freedom and $M$ is the number of Lyapunov exponents we would like to calculate. We typically take $M=2$ or $M=3$ but report only the largest exponent. The equation \eref{eq:linearised lyap} is then evolved together with $x(t)$ for a time $\Delta t$. After that time the vectors will start to align towards the directions related to the largest Lyapunov exponents. To counterbalance this alignment we apply a Gramm-Schmidt re-orthonormalization. The mean of the growth rate $\log(p_k(t))$ of the $k-$th orthonormal vector then determines the $k-$th finite-time Lyapunov exponent as
\begin{align}
    \lambda_k(t = N\Delta t) = \frac{1}{N \Delta t}\sum_{i=1}^N\log(p_k(i\Delta t)).
\end{align}
The Lyapunov exponents are then defined with $\lambda_k = \lim_{N\rightarrow\infty}\lambda_k(t=N\Delta t)$. In the main text, we plot the finite-time approximation to the largest Lyapunov exponent $\lambda_0(t)$. For integrable systems, this approximation converges to zero as $1/t$.

\section{Dynamical phase transition: additional results \label{app:dpt}}
\begin{figure*}[!h]
    \centering
    \includegraphics[width=0.45\textwidth]{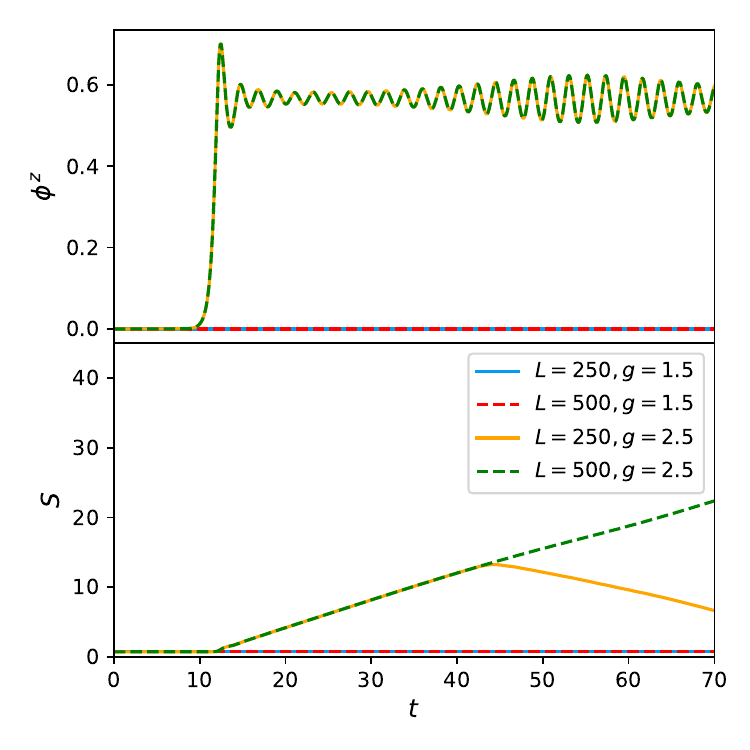}
    \includegraphics[width=0.45\textwidth]{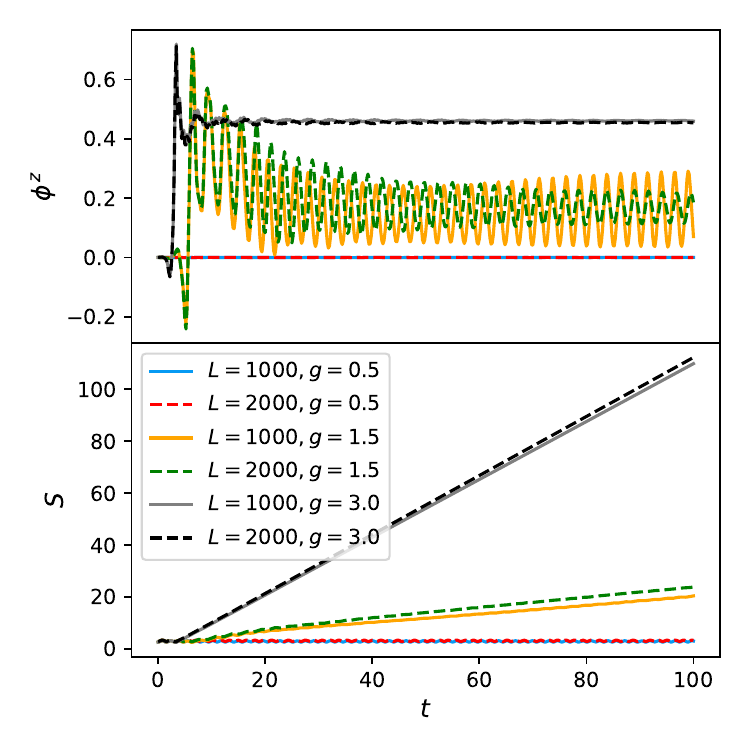}
    \caption{We show the systems size convergence of the finite-time Lyapunov exponents. The left panels correspond to the \texttt{quench I} and the right to the \texttt{quench II} discussed in the main text. We find converged exponents already for relatively small system sizes in the paramagnetic and standard ferromagnetic regions. In those cases, the finite-time Lyapunov exponents show the $1/t$ convergence to zero, which is typical for non-chaotic systems. In contrast, in the chaotic dynamical region \texttt{quench II} and $g=1.5$, we need larger system sizes to assess the order parameter in the thermodynamic limit.}
    \label{fig: global mf dynamical}
\end{figure*}

In this appendix, we provide additional plots regarding the system-size convergence in the XY model with infinite-range transverse interactions. In general, the convergence with system size and time is quick in the paramagnetic and standard ferromagnetic regions. However, it is much slower in the chaotic dynamical region. Slow convergence in the dynamical chaotic region is consistent with the observed chaotic behavior. In \fref{fig: global mf dynamical}, we show representative dynamics of the order parameter and the half-chain entanglement entropy for different system sizes. All quantities converge quickly with system size and time in the paramagnetic and ferromagnetic regions. In contrast, we need much larger systems and longer times to find convergence with system size and time in the chaotic dynamical phase. Even for system sizes $n=2000$, we still did not observe the convergence of the order parameter with the system size for all possible quenches.

\begin{figure*}[!h]
    \centering
    \includegraphics[width=0.35\textwidth]{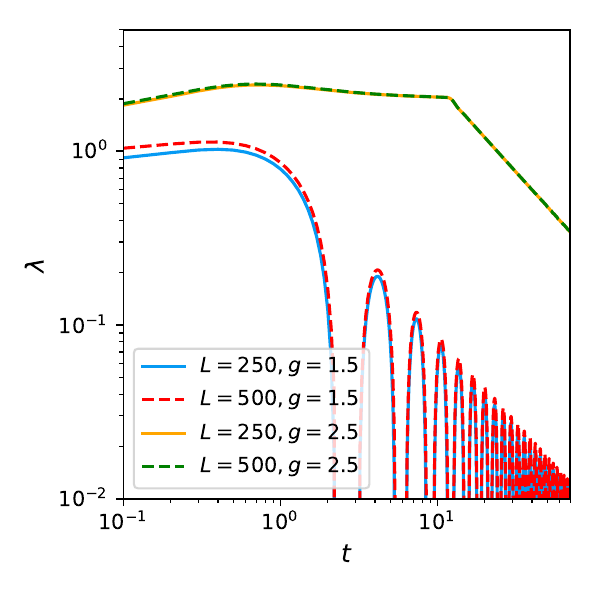}
    \includegraphics[width=0.35\textwidth]{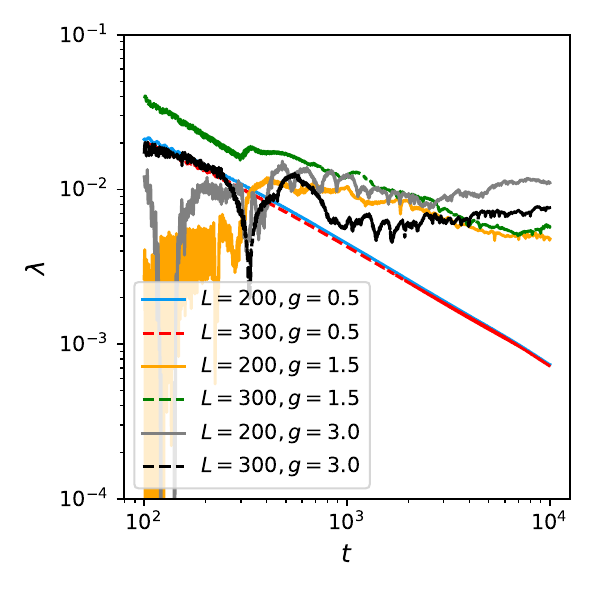}    
    \caption{We show the systems size convergence of the finite-time Lyapunov exponents. The left panels correspond to the \texttt{quench I} and the right to the \texttt{quench II} discussed in the main text. The finite-time Lyapunov exponents converge to $1/t$ behavior at relatively short times and small system sizes in the paramagnetic and standard ferromagnetic regions. In contrast, we need longer times to determine the finite-time Lyapunov exponents in the chaotic dynamical region, i.e., \texttt{quench II} and $g=1.5$.}
    \label{fig: global mf dynamical lyap}
\end{figure*}

In \fref{fig: global mf dynamical lyap}, we show the convergence of the finite-time Lyapunov exponents with time and system size. In the paramagnetic and the ferromagnetic regime, the finite-time Lyapunov exponents display the $1/t$ decay to zero, as expected for non-chaotic systems. In contrast, the finite-time Lyapunov exponents converge to a finite value in the chaotic dynamical phase. To find converged finite-time Lyapunov exponents in the chaotic dynamical phase, we had to increase the simulation times to $10^4/J$. Fortunately, the finite-time Lyapunov exponents did not change drastically with the system size. 

\begin{figure*}[!h]
    \centering
    \includegraphics[width=0.35\textwidth]{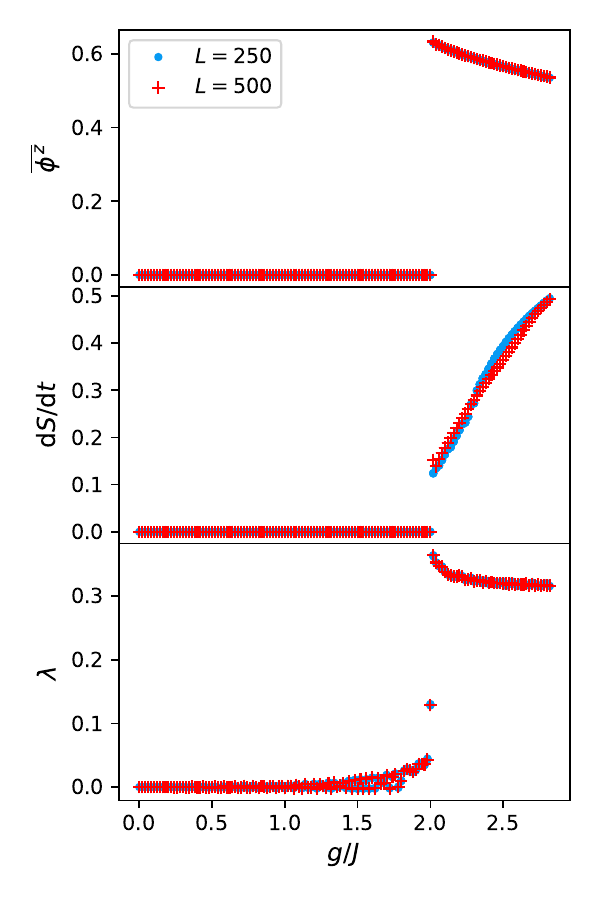}
    \includegraphics[width=0.35\textwidth]{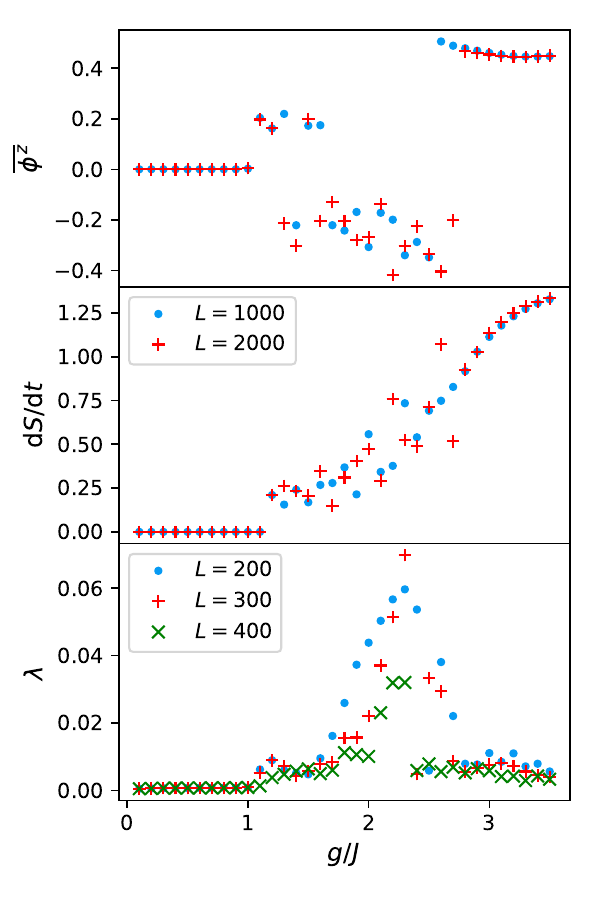}    
    \caption{We show the systems size convergence of the phase diagrams. The left panels correspond to the \texttt{quench I} and the right to the \texttt{quench II} discussed in the main text. As indicated in \fref{fig: global mf dynamical}, all quantities in the paramagnetic and standard ferromagnetic regions converge at relatively small system sizes. In contrast, we need large system sizes and long times to determine the order parameter and the finite-time Lyapunov exponents in the chaotic dynamical region \texttt{quench II}. Our order parameter and the entropy results are still not converged for accessible system sizes $L=2000$ and simulation times $T=100$. The finite-time Lyapunov exponents closer to the boundary of the chaotic dynamical phase seem to converge quickly with the system size. The finite-time Lyapunov exponents in the middle of the chaotic dynamical phase are still not converged with the system size even for the largest accessible system size $L=400$. To obtain the finite-time Lyapunov exponents, we used smaller system sizes and longer simulation times $T=10^4$.}
    \label{fig: global PD convergence}
\end{figure*}

Finally, in \fref{fig: global PD convergence}, we show the convergence of the phase diagram with the system size. In the \texttt{quench I}, we find a quick convergence with system size for all $g$. The finite-time Lyapunov exponent converges quickly with the system size and vanishes as $1/t$ in the paramagnetic and the ferromagnetic regions. In contrast, we need system sizes larger than $L=2000$ for the converged dynamical order parameter and entropy growth in the chaotic dynamical phase. At the boundary of the chaotic dynamical phase, we find converged finite-time Lyapunov exponents with system sizes $L=400$. In the middle of the chaotic region, we need larger system sizes, which were not accessible in our simulations. 

\newpage
\section{Local mean-field: additional results \label{app: local mf}}
\begin{figure}[!h]
    \centering
    \includegraphics[width=0.45\textwidth]{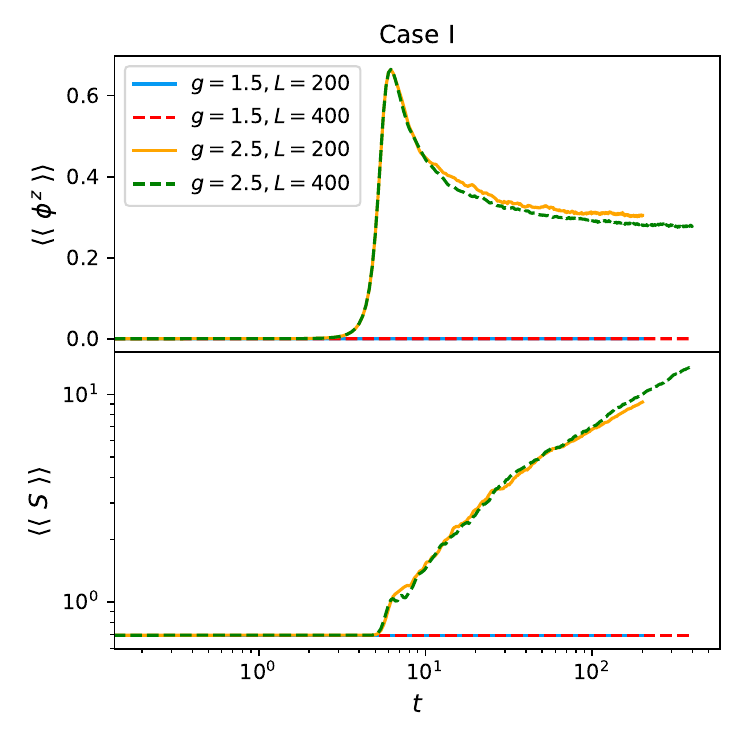}
    \includegraphics[width=0.45\textwidth]{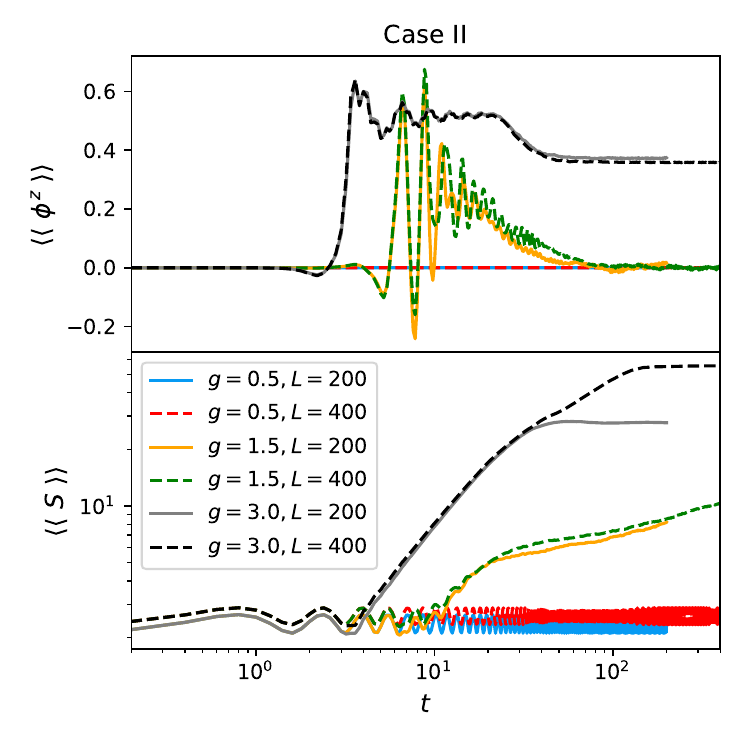}
    \caption{We show system size convergence of the time-dependent order parameter and the entropy. We show the average over ten disorder realizations. The entropy growth after a long time is slower than linear.}
    \label{fig: local MF dynamics}
\end{figure}

\begin{figure}[!h]
    \centering
    \includegraphics[width=0.5\textwidth]{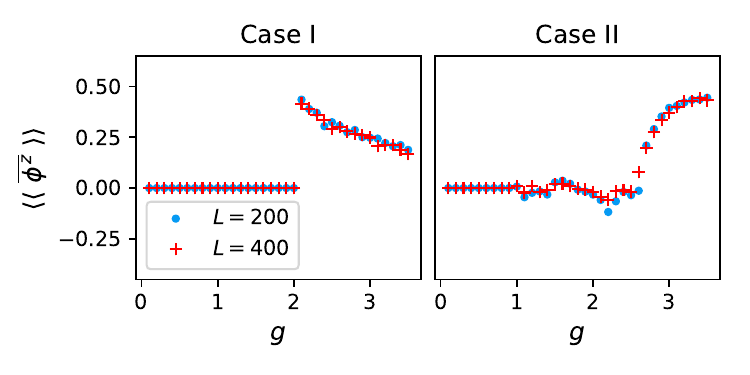}
    \caption{We show system size convergence of the dynamical order parameter. The left and the right panel correspond to the \texttt{quenches I} and \texttt{II}, respectively. The dots represent the average and the error bars the standard deviation over 10 initial disorder realizations with $h_i(t=0)\in[0,10^{-4}]$.  We find quick convergence of the dynamical order parameter with system size in all regimes.}
    \label{fig: local MF PD}
\end{figure}

We will now discuss the system size convergence of our local mean-field results. In \fref{fig: local MF dynamics}, we show the time dependence of the order parameter and the entropy for two system sizes. In contrast to the global mean-field model, we find quick convergence of the quantities in all regimes, a signature of a stabilized chaotic dynamical phase due to the (Anderson) localization. Similarly, the dynamical order parameter converges quickly with the system size (see \fref{fig: local MF PD}). 

\begin{figure}
    \centering
    \includegraphics[width=0.4\textwidth]{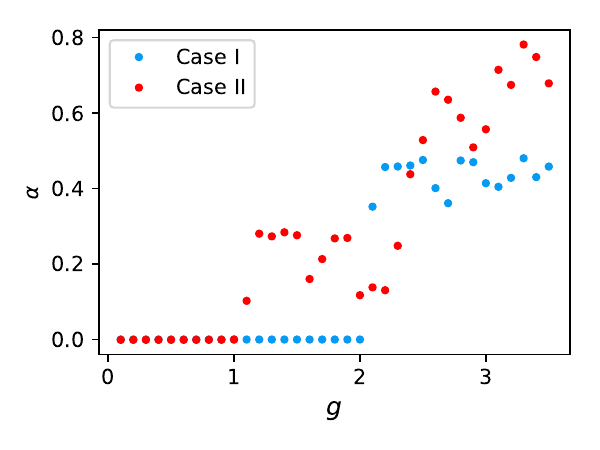}
    \caption{We show the fitted power law exponent of the long-time entropy growth $S(t)\propto t^\alpha$.}
    \label{fig: local s}
\end{figure}

Interestingly, the entropy growth slows down at long time. Our simulation times are not long enough to determine the asymptotic behavior, namely, if it is logarithmic or algebraic with some power $\alpha$. However, if we assume an algebraic long-time increase of the entanglement entropy, we find constant $\alpha=0.5$ in the \texttt{quench I} and (roughly linearly) increasing $\alpha$ with $g$ in the \texttt{quench II} (see \fref{fig: local s}).

\newpage
\section{Domain wall quench: system size convergence \label{app:dw}}
\begin{figure}[!htb]
    \centering
    \includegraphics[width=0.9\textwidth]{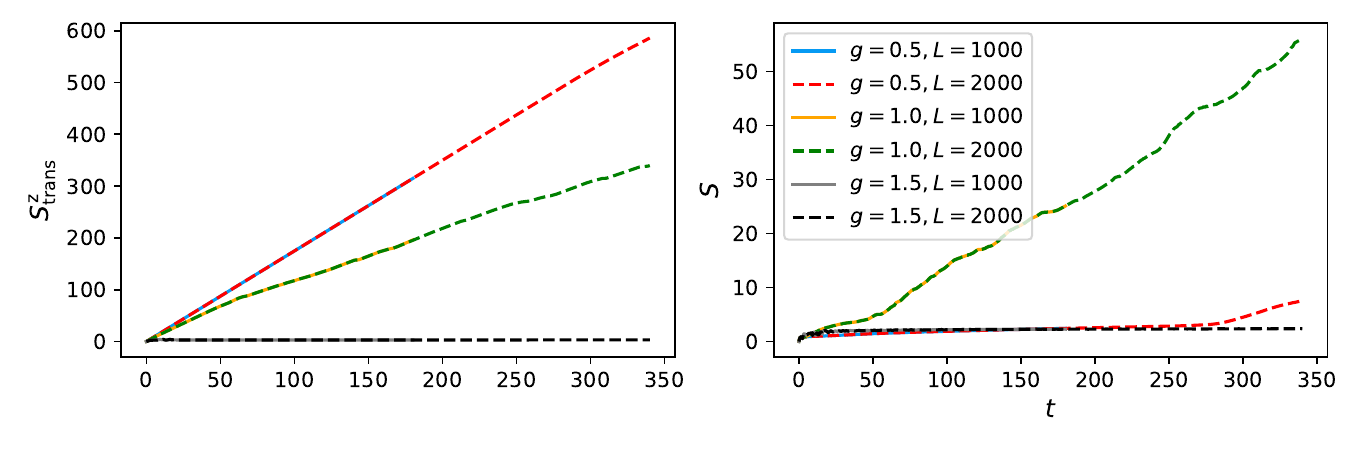}
    \caption{We show the convergence of the transferred magnetization and the entropy with system size. In all regimes, we find convergence with the system size $L=1000$. Due to logarithmic initial entropy growth, we need longer simulation times to find the long-time linear entropy growth (see examples with $g=0.5$). Therefore, we also need larger system sizes to avoid boundary effects.}
    \label{fig: local dw dynamics}
\end{figure}

In this appendix, we discuss the system size convergence of the inhomogeneous quench results. In \fref{fig: local dw dynamics}, we show the time evolution of the transferred transverse magnetization across the middle of the chain  $S^{\rm z}_{\rm trans}$. We find that our results converge with the system size $L=1000$. Interestingly, we find logarithmic entropy growth at short times up to a perturbative time, which grows as $1/g^2$. After that time, the entropy increases linearly. Therefore, the entropy growth is not converged with time for $g<0.5$.

\begin{figure}[!htb]
    \centering
    \includegraphics[width=0.9\textwidth]{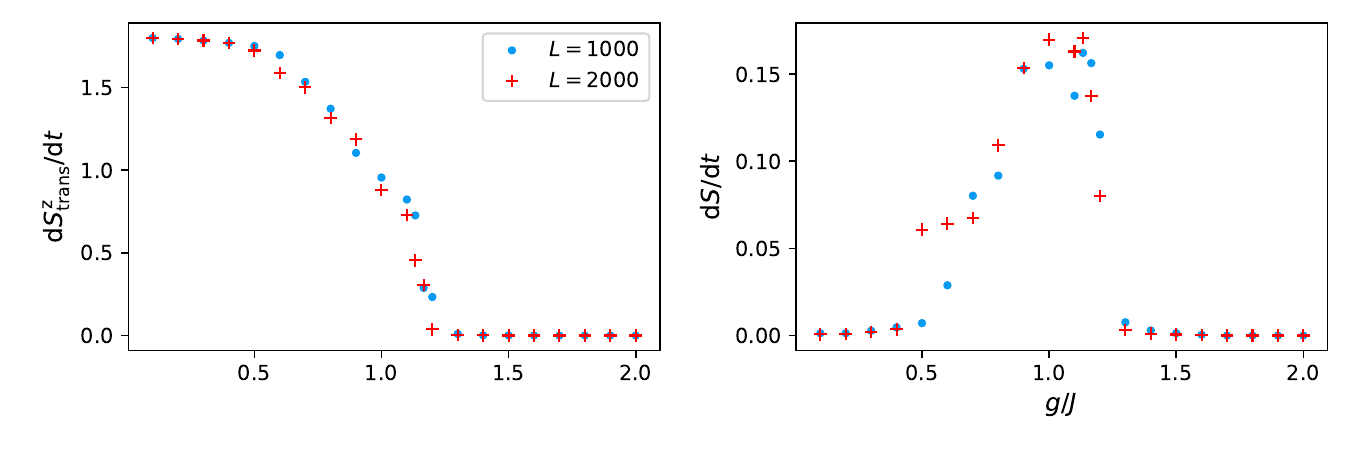}
    \caption{System size convergence of the phase diagram. The current ${\rm d}S^{\rm z}_{\rm trans}/{\rm d}t$ converges for the presented system sizes, while the entropy is not yet converged at small $g$ due to initial perturbative logarithmic growth.}
    \label{fig: local dw PD}
\end{figure}

In \fref{fig: local dw PD}, we show the system size convergence of the magnetization current and the entropy growth. The current converges in all regimes. In contrast, we need larger systems to assess the entropy growth at small $g$. Larger systems are necessary to avoid boundary effects due to longer simulation times. 

\begin{figure}[!htb]
    \centering
    \includegraphics[width=0.4\textwidth]{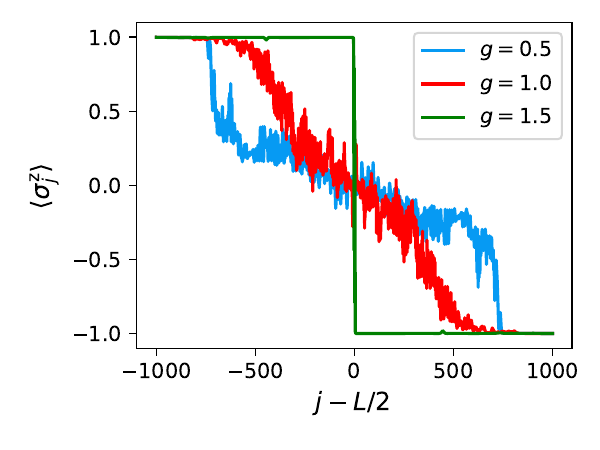}
    \caption{We show spin profiles in different regimes at $T=340$. In the insulating case, we have a step-like profile, which is very close to the initial profile. We find a site-dependent magnetization around a mean linear profile in the ballistic regime.}
    \label{fig: local dw profiles}
\end{figure}

Finally, in \fref{fig: local dw profiles}, we show the magnetization profiles at the final time $T=340$ for systems sizes $L=2000$ and typical values of the infinite-range interaction strength $g$. We find a step-like profile in the insulating regime and a linear profile in the ballistic regime. 


\end{document}